\documentstyle[12pt]{article}
\begin{document}
\newcommand{\beq}{\begin{equation}}
\newcommand{\eeq}{\end{equation}}
\newcommand{\beqa}{\begin{eqnarray}}
\newcommand{\eeqa}{\end{eqnarray}}
\newcommand{\bea}{\begin{eqnarray}}
\newcommand{\eea}{\end{eqnarray}}
\newcommand{\no}{\nonumber}
\newcommand{\grts}{\greaterthansquiggle}
\newcommand{\lets}{\lessthansquiggle}
\newcommand{\ul}{\underline}
\newcommand{\ol}{\overline}
\newcommand{\ra}{\rightarrow}
\newcommand{\Ra}{\Rightarrow}
\newcommand{\ve}{\varepsilon}
\newcommand{\vp}{\varphi}
\newcommand{\vt}{\vartheta}
\newcommand{\dg}{\dagger}
\newcommand{\wt}{\widetilde}
\newcommand{\wh}{\widehat}
\newcommand{\dfrac}{\displaystyle \frac}
\newcommand{\fsl}{\not\!}
\newcommand{\ben}{\begin{enumerate}}
\newcommand{\een}{\end{enumerate}}
\newcommand{\bfl}{\begin{flushleft}}
\newcommand{\efl}{\end{flushleft}}
\newcommand{\ba}{\begin{array}}
\newcommand{\ea}{\end{array}}
\newcommand{\btab}{\begin{tabular}}
\newcommand{\etab}{\end{tabular}}
\newcommand{\bit}{\begin{itemize}}
\newcommand{\eit}{\end{itemize}}
\newcommand{\be}{\begin{equation}}
\newcommand{\ee}{\end{equation}}
\newcommand{\bearr}{\begin{eqnarray}}
\newcommand{\eearr}{\end{eqnarray}}
\newcommand{\per}{\;\;.}
\newcommand{\pl}{PL}
\newcommand{\spi}{s_\pi}
\newcommand{\ql}{QL}
\newcommand{\qn}{QN}
\newcommand{\pn}{PN}
\newcommand{\qq}{Q^2}
\newcommand{\mee}{m_l^2}
\newcommand{\thp}{\theta_\pi}
\newcommand{\thl}{\theta_l}
\newcommand{\mtiny}[1]{{\mbox{\tiny #1}}}
\newcommand{\MS}{\mtiny{MS}}
\newcommand{\GeV}{\mbox{GeV}}
\newcommand{\MeV}{\mbox{MeV}}
\newcommand{\keV}{\mbox{keV}}
\newcommand{\ren}{\mtiny{ren}}
\newcommand{\kin}{\mtiny{kin}}
\newcommand{\hint}{\mtiny{int}}
\newcommand{\tot}{\mtiny{tot}}
\newcommand{\CHPT}{\mtiny{CHPT}}
\newcommand{\QED}{\mtiny{QED}}
\newcommand{\syst}{\mbox{syst.}}
\newcommand{\stat}{\mbox{stat.}}
\newcommand{\wave}{\mbox{wave}}
\newcommand{\co}{\; \; ,}
\newcommand{\nn}{\nonumber \\}
\newcommand{\fff}{\bar{f}}
\newcommand{\ffg}{\bar{g}}

\renewcommand{\theequation}{\arabic{section}.\arabic{equation}}
\renewcommand{\thetable}{\arabic{table}}

\begin{titlepage}

\def\mytoday#1{{ } \ifcase\month \or
 January\or February\or March\or April\or May\or June\or
 July\or August\or September\or October\or November\or December\fi
 \space \number\year}

\rightline{IPNO/TH 94-36}
\rightline{BUTP-94/11}
\rightline{ROM2F-94/19}
\rightline{hep-ph/9406211}

\vspace*{1cm}
\begin{center}

{\Large{\bf On the Pais--Treiman method}}\\

\indent

{\Large{\bf{ to measure $\pi\pi$
phase shifts in $K_{e4}$ decays}}}

\vspace{2cm}

{\bf{G. Colangelo}\footnote{Work supported in part by the INFN,
and by the HCM, EEC-Contract No. CHRX-CT920026 (EURODA$\Phi$NE)}}

\indent

{\sl Universit\"at Bern, Sidlerstrasse 5\\
CH$-$3012 Bern, Switzerland,\\
and\\
Dipartimento di Fisica,
Universit\`{a} di Roma II - "Tor Vergata"\\ I$-$00173 Roma, Italy.}

\indent

{\bf{M. Knecht}} and {\bf {J. Stern}}

\indent

{\sl Division de Physique Th\'eorique
\footnote{Unit\'e de Recherche des Universit\'es Paris XI et Paris VI
associ\'ee au CNRS.}, Institut de Physique Nucl\'eaire\\
F-91406 Orsay Cedex, France} \\

\indent

\vspace{2cm}

\end{center}

\indent

\begin{abstract}

\noindent
We evaluate theoretical uncertainties to the method of Pais and Treiman to
measure the $\pi\pi$ phase shifts in $K_{e4}$ decays. We find that they are
very small, below $1\%$.
\noindent

\end{abstract}
\end{titlepage}
\vfill \eject

\pagestyle{plain}
\clearpage
\setcounter{page}{1}
\setcounter{equation}{0}
\setcounter{subsection}{0}
\setcounter{table}{0}
\setcounter{figure}{0}

\section{Introduction}
The measurement of the $\pi\pi$ phase shifts near threshold is a key issue
of low energy hadronic physics. First, these phases enter the phenomenological
analysis of many different low energy scattering processes, weak decays and,
in particular, CP-violating K-decays. A model independent determination of
$\pi\pi$ phases would considerably improve our understanding of the
corresponding hadronic matrix elements. Furthermore, Chiral Perturbation
Theory (CHPT) \cite{revchir}, the systematic low energy expansion of QCD
amplitudes,
provides a link between the low energy $\pi\pi$ scattering data and the non
perturbative
chiral structure of the QCD ground state: The standard CHPT has a clean
prediction for these quantities \cite{glan,glnpb}, and would not be able
to explain
a large discrepancy with experimental data. On the other hand, the proposed
generalization \cite{SSF} of the usual low energy expansion,
called generalized
CHPT, expects a somewhat stronger I=0 S-wave $\pi\pi$ interaction. Within
the latter scheme, a discrepancy between the standard CHPT and experiment
would be interpreted as a manifestation of unusually low values of the
quark-antiquark condensate $\langle{\bar q}q\rangle$ and of the ratio of
strange
to  non strange current quark masses.

It is known since a long time that the main source of model independent
experimental information on low energy $\pi\pi$ phases comes from $K_{e4}$
decays. In these decays, a complete $\pi\pi$ phase-shift analysis could be
performed in principle, assuming nothing more than unitarity or the Watson
final state interaction theorem. However, such a complete procedure would
require a detailed amplitude analysis of $K_{e4}$ to be performed with
respect
to all five kinematical variables. To avoid this task, which is rather
problematic in practice, Pais and Treiman have proposed \cite{paistr}
to measure $\pi\pi$
phases in a much simpler way which, in addition to the Watson theorem,
assumes
that higher partial waves beyond the P-wave are small. A variant of the
Pais-Treiman method has already been used (together with additional
assumptions) in a 1977 high statistics experiment by Rosselet et al.
\cite{ross}, but the data obtained still show very large error bars.
Newly planned experiments like KLOE at DA$\Phi$NE,
will hopefully be able to reduce the errors sizeably enough to decide
between the theoretical alternatives mentioned above.
In view of this improvement on the experimental side, it is worth to check
what kind of uncertainties affect the Pais-Treiman method from the
theoretical point of view. This is the purpose of this note.

\setcounter{equation}{0}
\setcounter{subsection}{0}
\setcounter{table}{0}
\setcounter{figure}{0}
\section{Kinematics and form factors}

\label{kin}

We discuss the decay
         \bearr
          K^+(p) & \rightarrow & \pi^+(p_1) \;\pi^-(p_2) \;e^+(p_l) \; \nu_e
(p_{\nu})\co          \label{k1} \\
        \eearr
        and its charge conjugate mode.
 We do
        not consider isospin violating contributions and
        correspondingly set $m_u = m_d = {\hat m}$, $\alpha_\QED = 0$.

       The full kinematics of this decay requires five variables.
       We will use the
       ones introduced by Cabibbo and Maksymowicz \cite{cabmak}. It is
       convenient to
       consider three reference frames, namely the $K^+$ rest system
       $(\Sigma_K)$, the
       $\pi^+
       \pi^-$ center-of-mass system $(\Sigma_{2 \pi})$ and the
       $e^+\nu_e$ center-of-mass system $(\Sigma_{l \nu})$. Then
       the variables are
       $s_\pi$, the effective mass squared of the dipion system,
       $s_l$, the effective mass squared of the dilepton system,
       $\theta_\pi$, the angle of the $\pi^+$ in $\Sigma_{2\pi}$
       with respect to the dipion line of flight in $\Sigma_K$,
       $\theta_l$, the angle of the $l^+$ in $\Sigma_{l\nu}$
       with  respect to the dilepton line of flight in
       $\Sigma_K$, and
       $\phi$, the angle between the plane formed by the pions
       in
       $\Sigma_K$ and the corresponding plane formed by the dileptons.
      The range of the variables is
      \bearr
      4 M^2_\pi & \leq & s_\pi = (p_1+p_2)^2 \leq (M_K - m_l)^2 \co
      \nonumber \\
      m^2_l & \leq & s_l =(p_l+p_\nu)^2 \leq (M_K - \sqrt{s_\pi})^2\co
      \nonumber \\
      0 & \leq & \theta_\pi, \theta_l \leq \pi, 0 \leq \phi \leq 2
      \pi. \label{h6}
      \eearr

It is useful to furthermore introduce the following combinations of four
vectors
\be
P=p_1+p_2, \; \; Q=p_1-p_2,\; \; L=p_l+p_\nu .
\ee
Below we will also use the variables
\be \label{tu}
t=(p_1-p)^2 ,u=(p_2-p)^2 ,\nu=t-u.
\ee
These are related to $s_\pi,s_l$ and $\theta_\pi$ by
\bearr
t+u&=&2M_\pi^2 +M_K^2 +s_l -s_\pi \; , \nonumber \\
\nu&=& -2\sigma_\pi X \cos\theta_\pi \; ,
\eearr
where
\bearr
\sigma_\pi &=& (1-4M_\pi^2/s_\pi)^\frac{1}{2}\, ,  \nonumber \\
            X&=&\frac{1}{2}\lambda^{1/2}(M_K^2,s_\pi,s_l)\co \nn
\lambda (x,y,z) &=& x^2+y^2+z^2 -2(x y +x z +y z)\per
\eearr

      The matrix element is (we always neglect the mass of the electron
in what follows)
      \be
      T = \frac{G_F}{\sqrt{2}} V^\star_{us} \bar{u} (p_\nu) \gamma_\mu
      (1-\gamma_5) \nu(p_l) (V^\mu - A^\mu) \co
      \label{k11}
      \ee
      where
      \bearr
      I_\mu & = & < \pi^+ (p_1) \pi^- (p_2) \mbox{out}\mid
      I_\mu^{4-i5} (0) \mid K^+ (p)  >;\; I = V,A \co
      \nonumber \\
      V_\mu & = & - \frac{H}{M^3_K} \epsilon_{\mu \nu \rho \sigma} L^\nu
      P^\rho Q^\sigma \co
      \nonumber \\
      A_\mu & = & -i\frac{1}{M_K} \left [ P_\mu F +
      Q_\mu G + L_\mu R \right ] \co
      \label{k12}
      \eearr
and $\epsilon_{0123}=1$.

      The form factors $F,G,R$ and $H$ are
analytic functions of the variables $s_\pi, t$ and $u$.
      The partial decay rate for (\ref{k1}) can be expressed as

 \bearr\label{k14}
      d \Gamma_5 &=& G^2_F \mid V_{us} \mid^2  N(s_\pi, s_l) J_5
      (s_\pi, s_l, \theta_\pi, \theta_l, \phi) ds_\pi ds_l d (\cos
      \theta_\pi) d(\cos \theta_l) d\phi \co \nn
      N(s_\pi, s_l) &=&  \sigma_\pi X
      /(2^{13} \pi^6 M_K^5) \co
      \eearr
      where
\bearr
J_5 &=&2\left[ I_1 + I_2 \cos 2 \theta_l + I_3 \sin^2 \theta_l
 \cdot \cos 2
\phi + I_4 \sin 2 \theta_l \cdot \cos \phi \right. \nonumber \\
&+& \left. I_5 \sin \theta_l \cdot
\cos \phi
+  I_6 \cos \theta_l + I_7 \sin \theta_l \cdot \sin \phi + I_8 \sin 2
\theta_l \cdot \sin \phi \right. \nonumber \\
 &+& \left. I_9 \sin^2 \theta_l \cdot \sin 2 \phi \right] \co
\nonumber
      \eearr
with
      \bearr
I_1 &=& \frac{1}{4}\left\{ |F_1|^2  + \frac{3}{2}
\left(|F_2|^2 +
|F_3|^2 \right) \sin^2 \thp   \right\}
\co \nn
I_2 &=& - \frac{1}{4} \left\{ |F_1|^2 - \frac{1}{2}
\left( |F_2|^2 + |F_3|^2 \right)  \sin^2 \thp  \right\}
\co \nn
I_3 &=& - \frac{1}{4} \left\{ |F_2|^2 - |F_3|^2 \right\}\sin^2 \thp
\co \nn
 I_4 &=& \frac{1}{2}\mbox{ Re} (F_1^* F_2)  \sin \thp
\co \nn
I_5 &=& - \mbox{ Re} (F_1^* F_3) \sin \thp
\co \nn
I_6 &=&
- \mbox{ Re} (F_2^* F_3)\sin^2 \thp
\co \nn
I_7 &=& - \mbox{ Im} (F_1^* F_2) \sin \thp
\co \nn
I_8 &=& \frac{1}{2} \mbox{ Im} (F_1^* F_3) \sin \thp
\co \nn
I_9 &=& -\frac{1}{2} \mbox{ Im} (F_2^* F_3) \sin^2 \thp \co
      \eearr
and
      \bearr\label{ki30}
 F_1& =& X \cdot F + \sigma_\pi (P L) \cos \thp \cdot G
\co \nn
 F_2& =& \sigma_\pi \left( s_\pi s_l \right)^{1/2} G
\co \nn
 F_3& =&  \sigma_\pi X \left( s_\pi s_l \right)^{1/2}  \frac{H}{M_K^2}
\co \per
      \eearr
      The definition of $F_1$, $F_2$ and $F_3$ corresponds
      to the combinations used by Pais and Treiman \cite{paistr}.
      The form factors $I_1, \ldots, I_9$
      agree with the expressions given in \cite{paistr} if one neglects the
      mass of the electron.

      The form factors may be written in a partial wave expansion in
      the variable $\theta_\pi$.
      Suppressing isospin indices, one has for $F$ and $G$ \cite{berends}
      \bearr
      F &= & \sum^{\infty}_{l=0} P_l (\cos \theta_\pi) f_l e^{i\delta_l}
      - \frac{\sigma_\pi PL}{X} \cos
      \theta_\pi G \co
      \nonumber\\
      G & = & \sum^{\infty}_{l=1} P_l' (\cos \theta_\pi) g_l e^{i\delta_l} \co
      \label{i4}
      \eearr
      where
      \be
      P_l'(z) = \frac{d}{dz} P_l(z) \; \; .
      \label{i10}
      \ee

      The partial wave amplitudes $f_l$ and $g_l$ depend on
      $s_\pi$ and $s_l$ and are real in the physical region of $K_{l4}$
      decay (in our overall phase convention).
      The phases $\delta_l$ coincide with the phase shifts
      in elastic $\pi \pi$ scattering.
      Accordingly, the form factors $F_1$ and $F_2$ have the expansions:
\bearr
      F_1 &= & X \sum^{\infty}_{l=0} P_l (\cos \theta_\pi) f_l
      e^{i\delta_l}\co
      \nn
      -\sin\thp F_2 & = & \sigma_\pi (\spi s_l)^{1/2}
      \sum^{\infty}_{l=1} P_l^{(1)} (\cos \theta_\pi) g_l e^{i\delta_l} \co
      \label{i6}
\eearr
where $P^{(m)}_l$ are the associated Legendre functions.

\setcounter{equation}{0}
\setcounter{subsection}{0}
\setcounter{table}{0}
\setcounter{figure}{0}

\section{Evaluation of the corrections}
The method suggested by Pais and Treiman to measure the $\pi\pi$ phase shifts
in $K_{e4}$ decays is very simple and clean. It is based on the observation
that the dependence of the differential decay rate on two of the five
variables can be worked out analytically under very few general assumptions,
as we have seen in the previous section.

In case one neglects all the waves higher than $S$ and $P$, $I_4$ and $I_7$
show a very simple dependence on the phases $\delta_0$ and $\delta_1$.
[The same
dependence appears in $I_5$ and $I_8$ which, however, contain the anomalous
form factor $H$, and are kinematically much more suppressed.] Their integral
over $\cos \thp$ is:

\bearr
{\bar I_4}&\equiv&\int_{-1}^1 d\cos\thp I_4= \frac{\pi}{4}
X \sigma_\pi (\spi s_l)^{1/2} f_0 g_1 \cos(\delta_0-\delta_1)
\co \nn
{\bar I_7}&\equiv&\int_{-1}^1 d\cos\thp I_7= \frac{\pi}{2}
X \sigma_\pi (\spi s_l)^{1/2} f_0 g_1 \sin(\delta_0-\delta_1)
\per
\eearr
By measuring the ratio ${\bar I_7}/2{\bar I_4}$ one has then direct access to
$\tan(\delta_0-\delta_1)$.
To experimentally select $I_7$ and $I_4$ one could use different methods
(fit the distribution in $\cos \thl$ and $\phi$, measure asymmetries, measure
moments \ldots). Choosing for example to measure the appropriate moments,
one would have, all in all, to integrate the distribution over four of the
five
variables with the weights $\sin \phi$, then $\cos \phi \cos \thl$, and then
take the ratio. The measurement of $\tan(\delta_0-\delta_1)$ would be as
simple as that.

While on the experimental side there might be detector-dependent problems
that could force experimentalists to adapt the Pais-Treiman method to their
own peculiar situation (see for example Ref. \cite{ross}), it is clear from
the theoretical point of view that the beauty of the method is based on
neglecting
the higher waves. It is the main purpose of this note to try to estimate
corrections coming from higher waves to the Pais-Treiman formula:

\beq \label{PT}
\frac{{\bar I_7}}{2{\bar I_4}} = \tan(\delta_0-\delta_1) \per
\eeq
For that purpose we shall use the form factors predicted by CHPT at the one
loop level.

We start by applying the chiral power counting to the partial waves of the
form factors. In general, for any partial wave we may write an expansion
in the following way:
\be
X_l=\frac{M_K}{\sqrt{2}F_\pi}\left\{X_l^{(0)} + X_l^{(2)} + X_l^{(4)} +\cdots
\right\}\;;\;
X=f,g\co
\label{in2}
\ee
where the upper index stands for powers of energies or meson masses. From
published calculations of $F$ and $G$ to one-loop \cite{bijnenskl4,riggen}
we may easily get that
only three partial waves start at order $O(E^0)$, while all the others start
at order $O(E^2)$:
\bearr
f_0&=&\frac{M_K}{\sqrt{2}F_\pi}\left\{1 + f_0^{(2)} + O(E^4)
\right\}\;;\; \nn
f_1&=& \frac{M_K}{\sqrt{2}F_\pi}\left\{ \frac{\sigma_\pi PL}{X} + f_1^{(2)}
+ O(E^4)
\right\}\;;\; \nn
g_1&=&\frac{M_K}{\sqrt{2}F_\pi}\left\{1 + g_1^{(2)} + O(E^4)
\right\}\;;\; \nn
f_l&=&\frac{M_K}{\sqrt{2}F_\pi}\left\{f_l^{(2)} + O(E^4)
\right\}\;; l\geq 2 \; \nn
g_k&=&\frac{M_K}{\sqrt{2}F_\pi}\left\{g_k^{(2)} + O(E^4)
\right\}\;; k\geq 2 \; \per
\eearr

Using this chiral power counting is very easy to give a "corrected"
Pais-Treiman formula, which is accurate up to and including order $O(E^2)$:

\beq \label{cordef}
\frac{{\bar I_7}}{2{\bar I_4}}\equiv \tan(\delta_0-\delta_1)\left\{1
+ \Delta^{(2)} +O(E^4)\right\} \co
\eeq
\bearr
\Delta^{(2)}&=&
 \left( \frac{\sin \delta_0}{\sin(\delta_0-\delta_1)}-
\frac{\cos \delta_0}{\cos(\delta_0-\delta_1)}\right)
\sum_{k=1}^{\infty}A_{0\, 2k+1} g^{(2)}_{2k+1}  \nn
&+&  \left(\frac{\sin \delta_1}{\sin(\delta_0-\delta_1)}-
\frac{\cos \delta_1}{\cos(\delta_0-\delta_1)}\right)
\frac{\sigma_\pi PL}{X}
\sum_{k=1}^{\infty}A_{1\, 2k} g^{(2)}_{2k} \nn
&-& \left(\frac{\sin \delta_1}{\sin(\delta_0-\delta_1)}+
\frac{\cos \delta_1}{\cos(\delta_0-\delta_1)}\right)
\sum_{l=1}^{\infty}A_{2l\, 1} f^{(2)}_{2l}\co
\eearr
where
\beq
A_{l\,k} = \frac{2}{\pi} \int^{1}_{-1}d\cos\thp P_l(\cos\thp)
P^{(1)}_k(\cos\thp)
\co
\eeq
and where all the phases of waves higher than the $P$ have been put to zero
(this is again consistent at the order at which we are working).
The ratios of sines and cosines start at order $O(E^0)$, and will have also
contributions of higher order, that we neglect at this level of accuracy.
We may thus use the leading order CHPT expressions:
\bearr
\delta_0&=&\frac{1}{32\pi F^2_\pi} \sigma_\pi \left(2 \spi -M_\pi^2
+5\epsilon M_\pi^2 \right)
+O(E^4)\\
\delta_1&=&\frac{1}{96\pi F^2_\pi} \sigma_\pi^3 \spi +O(E^4) \per
\eearr
The standard CHPT predicts $\epsilon =0$, whereas in generalized CHPT
$\epsilon$ is an arbitrary parameter, $0\le\epsilon \le1$, related to
the quark mass ratio $r= m_s /{\hat m}$ \cite{SSF}. Eq. (3.6) then becomes

\bearr \label{corr1}
\Delta^{(2)}&=&
\left(\frac{\spi-4M_\pi^2}{5\spi+M_\pi^2 + 15\epsilon M_\pi^2}
\right) \sum_{k=1}^{\infty}A_{0\, 2k+1} g^{(2)}_{2k+1} \nn
&-&  \left(\frac{4\spi+5M_\pi^2 +15\epsilon M_\pi^2}{5\spi+M_\pi^2
+15\epsilon M_\pi^2}
\right)\frac{\sigma_\pi PL}{X}
\sum_{k=1}^{\infty}A_{1\, 2k} g^{(2)}_{2k} \nn
&-& \left(\frac{6\spi-3M_\pi^2 +15\epsilon M_\pi^2}{5\spi+M_\pi^2
+15\epsilon M_\pi^2}
\right) \sum_{l=1}^{\infty}A_{2l\, 1} f^{(2)}_{2l} \per
\eearr

Before evaluating numerically the correction we would like to stress that
the low energy expansion has been performed only inside the braces of eq.
(\ref{cordef}).

The explicit calculation shows that both in standard and in generalized CHPT
$\Delta^{(2)}$ is numerically very small
over the whole phase space. It barely reaches $0.5$ \%. The
dependence on $s_l$ is rather weak, and there is no difficulty in repeating
the same analysis for the ratio of ${\bar I_7}$ and $2{\bar I_4}$ not
treated as functions
of $s_l$ but averaged over it (as suggested by
Pais and Treiman). At this order in the low energy expansion the
correction turns out to be just $\langle \Delta^{(2)} \rangle$, where with
the symbol $\langle \rangle$ we mean the following average:
\beq
\langle A \rangle \equiv \frac{1}{\int_{s_l^{min}}^{s_l^{max}}
ds_l X^2 \sqrt{s_l}}
\int_{s_l^{min}}^{s_l^{max}} ds_l X^2 \sqrt{s_l} A(s_l)
\eeq
The plot of $\langle \Delta^{(2)} \rangle$ as a function of
$\spi$ is shown in fig. 1.

Since the corrections are so small, we would like to understand whether this
is just a typical size of higher waves, or an effect depending on some
miraculous cancellation.

Since $F$ and $G$ are dimensionless functions of $\spi$, $s_l$ and $\nu$,
and depend on $\cos \theta_\pi$ only through $\nu$, their higher waves
projections are suppressed by powers of the kinematical factor:
\beq\label{kinet}
\frac{{\sigma}_{\pi} X}{M_K^2} = 2\frac{p_{\pi}p_l}{M_K^2}\le0.15
\ee
where $p_{\pi}$ and $p_l$ stand for the momenta of individual pions and
of the dilepton, respectively, in the dipion center of mass frame. The
bound in (\ref{kinet}) holds over the whole phase space.

The dominant contribution to the
sum $\Delta^{(2)}$, eq. (\ref{corr1}), comes from the D-waves
$g^{(2)}_{2}$ and
$f^{(2)}_{2}$.
The latter contains contributions both from the $F$ and $G$ form factors:
\beq
f_2^{(2)}=f_d^{(2)}+2\frac{\sigma_\pi PL}{X} g_2^{(2)} \co
\eeq
where $f_d^{(2)}$ stands for the D-wave of $F$ alone.
Making explicit the factors (\ref{kinet}), we may define:
\bearr
f_d^{(2)}&\equiv&\left(\frac{\sigma_\pi X}{M_K^2}\right)^2
\widetilde{f_d}^{(2)} \co \nn
g^{(2)}_2 &\equiv&\frac{\sigma_\pi X}{M_K^2} \widetilde{g_2}^{(2)}
\co
\eearr
where $\widetilde{f_d}^{(2)}$ and $\widetilde{g_2}^{(2)}$ are expected to be
smooth functions of $s_\pi$ and of $s_l$.
Since D-waves are a loop effect, we may guess that $\widetilde{f_d}^{(2)}$
and $\widetilde{g_2}^{(2)}$ will be of the order $M_K^2/(16 \pi^2 F_\pi^2)$.
An explicit calculation shows that this is
indeed the case and, for a possible later  use, we may even write the
following simplified expressions for them: Standard CHPT gives
\bearr \label{numbers}
\widetilde{f_d}^{(2)}&=&\frac{M_K^2}{16\pi^2 F_\pi^2} (-0.28)
\left[1+0.12 \left(\frac{s_l}{4M_\pi^2}-q^2\right)\right] \co \nn
\widetilde{g_2}^{(2)}&=&\frac{M_K^2}{16\pi^2 F_\pi^2} (-0.62)
\left[1+0.08\left(\frac{s_l}{4M_\pi^2}-q^2\right)\right] \co
\eearr
where

\beq
q^2=(\spi-4M_\pi^2)/4M_\pi^2 \per
\eeq
The above evaluation of $\widetilde{f_d}^{(2)}$ is parameter free.
On the other hand, a substancial part of the contribution to
$\widetilde{g_2}^{(2)}$ comes from the low energy constant $L_3 + 4L_2$,
whose determination brings in some uncertainty \cite{bicolga}:
The corresponding error in (3.15) is however smaller than 50\%.
Finally, $\Delta^{(2)}$ can be expressed in terms of the smooth
functions $\widetilde{f_d}^{(2)}$ and $\widetilde{g_2}^{(2)}$:
\bearr \label{corr2}
\Delta^{(2)}&=& \frac{3}{8}
\left(\frac{2\spi-M_\pi^2 +5\epsilon M_\pi^2}{5\spi+M_\pi^2 + 15\epsilon
M_\pi^2}
\right) \left( \frac{\sigma_\pi X}{M_K^2}\right)^2 \widetilde{f_d}^{(2)} \nn
&-& \frac{3}{2} \left(\frac{\spi+3M_\pi^2 +5\epsilon M_\pi^2}{5\spi+M_\pi^2 +
15\epsilon M_\pi^2}
\right)\frac{\sigma_{\pi}^2 (PL)}{M_K^2} \widetilde{g_2}^{(2)} + \cdots
\eearr
where the ellipsis stands for $l>2$ waves.
This expression gives clearly account of the size of the corrections. The
second term is dominant with respect to the first one, and its typical scale
is $\sigma_{\pi}^2 (PL)/(16\pi^2 F_{\pi}^2)$, which, after averaging over
$s_l$,  is at most $1.6\%$ over the
whole phase space. The remaining coefficients reduce this number by a factor
$4$. Moreover, we have calculated the dependence of $\widetilde{f_d}^{(2)}$
and $\widetilde{g_2}^{(2)}$ on $m_s/{\hat m}$, in generalized
CHPT\footnote{The generalized CHPT one loop $K_{l4}$ form factors will
be given elsewhere \cite{knest} }. We have found that
$\widetilde{f_d}^{(2)}$ is affected by no
more than 10\%, whereas the form factor $\widetilde{g_2}^{(2)}$  may be
modified by at most a factor one half, further reducing
$\langle \Delta^{(2)}\rangle$.

Higher orders will certainly modify the above numbers, but not the orders of
magnitude, which are essentially given by the kinematics. In a sense, the
limited effect of the uncertainties in the values of the low energy
constants $L_i$ may be considered as an indication in this direction.
Furthermore, since the generalized CHPT differs from the standard CHPT
by a different ordering in
the perturbative expansion (that is, at every finite order it includes
terms which in the standard scheme are relegated to higher orders), the weak
dependence of $\widetilde{f_d}^{(2)}$ and of $\widetilde{g_2}^{(2)}$ on
$m_s/{\hat m}$ is a further indication that it is very unlikely that higher
orders would overwhelm the strong kinematical suppression of higher waves.

Our conclusion is then that the Pais-Treiman formula
(\ref{PT})
 is free of corrections up to the percent level, over the whole accessible
range of energy.

\indent

\indent

\noindent {\bf Acknowledgements}

One of us (G.C.) would like to thank J\"urg Gasser for pointing
out to him the
relevance of the issue, and the Division de Physique Th\'eorique de
l'Institut de Physique Nucl\'eaire
d`Orsay for a very kind hospitality.
It is a pleasure to thank J. Bijnens, J. Gasser, E. Pallante and
R. Petronzio for very informative discussions.

\indent

\indent

\indent

\noindent {\Large{\bf Figure Caption}}

\indent

Fig. 1: $\langle \Delta^{(2)} \rangle$, as a function of $s_\pi$. The plot
refers to the standard CHPT case: it is calculated using eq. (\ref{corr1})
with $\epsilon=0$ and the $F$ and $G$ form factors given by
\cite{bijnenskl4,riggen}, with the central values $L_2 = 1.35\times 10^{-3}$,
$L_3 = -3.5\times 10^{-3}$ of \cite{bicolga}. The same curve can be obtained
using eq. (\ref{corr2}) with $\epsilon=0$, and (\ref{numbers}).


\begin{thebibliography}{99}

\newcommand{\PL}[3]{{Phys. Lett.}        {#1} {(19#2)} {#3}}
\newcommand{\PRL}[3]{{Phys. Rev. Lett.} {#1} {(19#2)} {#3}}
\newcommand{\PR}[3]{{Phys. Rev.}        {#1} {(19#2)} {#3}}
\newcommand{\NP}[3]{{Nucl. Phys.}        {#1} {(19#2)} {#3}}

\bibitem{revchir}
For recent reviews on CHPT see e.g. \\
H. Leutwyler, in: Proc. XXVI
Int. Conf. on High Energy Physics, Dallas, 1992, edited by J.R. Sanford,
AIP Conf. Proc. No. 272 (AIP, New York, 1993) p. 185;\\
U.G. Mei\ss ner, Rep. Prog. Phys. 56 (1993) 903; \\
A. Pich,
Lectures given at the
V Mexican School of Particles and Fields,
Gu\-ana\-juato, M\'exico, December 1992, preprint CERN-Th.6978/93
(hep-ph/9308351);\\
G. Ecker,
Lectures given at the
$6^{\rm th}$ Indian--Summer School on Intermediate Energy Physics
Interaction in Hadronic Systems
Prague, August 25 - 31, 1993, to appear in the Proceedings (Czech. J. Phys.),
preprint UWThPh -1993-31 (hep-ph/9309268).





\bibitem{glan}
 J. Gasser and H. Leutwyler, Ann. Phys. (N.Y.) 158 (1984) 142.

\bibitem{glnpb}
J. Gasser and H. Leutwyler, Nucl. Phys. B250 (1985) 465.



\bibitem{SSF} J. Stern, H. Sazdjian, N.H. Fuchs, \PR{D47}{93}{3814}


\bibitem{paistr}
A. Pais and S.B. Treiman, Phys. Rev. 168 (1968) 1858.

\bibitem{ross}
L. Rosselet et al., Phys. Rev. D15 (1977) 574.


\bibitem{cabmak}
N. Cabibbo and A. Maksymowicz, Phys. Rev. B137 (1965) B438;
 Phys. Rev. 168 (1968) 1926 E.


\bibitem{berends}
F.A. Berends, A. Donnachie and G.C. Oades, Phys. Lett. 26B (1967) 109;
Phys. Rev. 171 (1968) 1457.


\bibitem{bijnenskl4}
J. Bijnens, Nucl. Phys. B337 (1990) 635.


\bibitem{riggen}
C. Riggenbach,  J. Gasser, J.F. Donoghue and B.R. Holstein, Phys. Rev.
D43  (1991) 127.



\bibitem{bicolga}
J. Bijnens, G. Colangelo and J. Gasser, $K_{l4}$-decays beyond one loop,
preprint BUTP-94/4 and ROM2F 94/05 (hep-ph/9403390).

\bibitem{knest}
M. Knecht and J. Stern, to appear.

\end{thebibliography}
\end{document}